\documentclass{osa-article}

\journal{osajournal}
\articletype{Research Article}

\usepackage{float}
\usepackage{color, soul}
\usepackage{siunitx}

\begin{document}

\title{Cancellation of photothermally induced instability in an optical resonator}

\author{Jiayi Qin$^1$, Giovanni Guccione$^1$, Jinyong Ma$^{2}$, Chenyue Gu$^1$, Ruvi Lecamwasam$^{1,3}$, Ben C.\ Buchler$^1$, Ping Koy Lam\authormark{*}$^{1,4}$}

\address{$^1$Centre for Quantum Computation and Communication Technology, Department of Quantum Science and Technology, Research School of Physics, The Australian National University, Canberra ACT 2601, Australia.}

\address{$^2$ ARC Centre of Excellence for Transformative Meta-Optical Systems, Research School of Physics, The Australian National University, Canberra ACT 2601, Australia.}

\address{$^3$ Quantum Machines Unit, Okinawa Institute of Science and Technology Graduate University, Okinawa 904-0495, Japan}

\address{$^4$ Institute of Materials Research and Engineering, Agency for Science Technology and Research (A*STAR), 2 Fusionopolis Way, 08-03 Innovis 138634, Singapore}

\email{\authormark{*}ping.lam@anu.edu.au} 



\begin{abstract*}
\textbf{Optical systems are often subject to parametric instability caused by the delayed response of the optical field to the system dynamics. In some cases, parasitic photothermal effects aggravate the instability by adding new interaction dynamics. This may lead to the possible insurgence or amplification of parametric gain that can further destabilize the system. In this paper, we show that the photothermal properties of an optomechanical cavity can be modified to mitigate or even completely cancel out optomechanical instabilty. By inverting the sign of the photothermal interaction to let it cooperate with radiation pressure, we achieve control of the system dynamics to be fully balanced around a stable equilibrium point. Our study provides a new feedback solution for optical control and precise metrological applications, specifically in high-sensitivity resonating systems that are particularly susceptible to parasitic photothermal effects, such as our test case of a macroscopic optical levitation setup. This passive stabilization technique is beneficial for improving system performance limited by photothermal dynamics in broad areas of optics, optomechanics, photonics and laser technologies.} \copyright 2022 Optica Publishing Group under the terms of the \href{https://opg.optica.org/library/license_v2.cfm#VOR-OA}{Open Access Publishing Agreement}

\noindent
\href{https://doi.org/10.1364/OPTICA.457328}{https://doi.org/10.1364/OPTICA.457328}
\end{abstract*}

\section{Introduction}
Light is a powerful medium to enable the most precise metrological measurements~\cite{Abb16, Min20}, interact with the smallest components~\cite{Teu11}, and prepare important resources for quantum information~\cite{Sch17, Evg19, Guc20}. The field of optomechanics~\cite{Asp14} uses the radiation pressure interaction between light and mechanical elements for both applied and fundamental applications such as in high-precision atomic force microscopy~\cite{Liu12}, quantum non-demolition measurements~\cite{Lec15}, and cooling of mechanical oscillators to ground states~\cite{Chan11}. Optical tweezers~\cite{Ashkin87, Anderegg19, Garces02, Li11}, commonly applied in different research fields as a convenient tool for the manipulation of microscopic objects, are a noteworthy example of optomechanical control. The refinement of this type of optical interaction has made it possible to reach the motional ground state of levitated nanoparticles even at room temperature conditions~\cite{Del20}.

A striking form of optomechanical control is the optical spring. An optical spring effect can occur in an optical cavity, where the standing wave of optical field is modulated by the position of cavity mirror so as to provide restoring and anti-restoring forces on the mirror. This optically induced stiffness can be used for optical control of the oscillator's trap~\cite{Corbitt07, Altin17}, for optical dilution of mechanical dissipation mechanisms~\cite{Corbitt2007}, and for macroscopic objects in the quantum regime~\cite{Guc13, Singh10}. The optical spring has many advantages over conventional spring systems in regards of its tunability, robustness to mechanical dissipation, and versatility for real-time information readouts. 

With the development of new and more extreme cavity optomechanical systems at all scales, increased optical power density and ever smaller physical dimensions can lead to some deleterious side effects. Photothermal interaction is one such side effect. Heating induced by optical absorption results in mechanical driving of the system. This driving can combine with parasitic mechanical modes and lead to parametric instabilities that inevitably affect the performance of the instrument~\cite{Zhao15}. This limitation has been observed to be dominant in micromechanical systems~\cite{Rok05} as well as high-power systems, including levitated sensor detectors (LSD)~\cite{aggarwal2022searching} and optical levitation configurations~\cite{Ma20} . In these systems, the photothermal effects become a source of noise and instability~\cite{Evans08, Cerdonio01}. Even large-scale gravitational wave interferometers such as LIGO and Virgo --- among the best controlled experimental environments ever built --- are not immune to parasitic photothermal effects~\cite{Rosa02, Evan15}. Recent studies has also showed growing interest in the use of photothermal dynamics at versatile applications such as high power and pulsed lasers~\cite{stubenvoll2013photothermal}, monolithic and micro-resonators, waveguides, and fibre cavities~\cite{jiang2020optothermal, qiu2022dissipative}; for example, the demonstrations of optothermal spectroscopy and optical nonreciprocity in optical whispering-gallery-mode microresonators require the assistance of photothermal bistability~\cite{jiang2020optothermal}. Photothermal feedback was proposed for future dissipation engineering in both optical and superconducting microwave cavities~\cite{qiu2022dissipative}.

Photothermal effects arise from the absorption of light by the object, followed by thermal expansion or a thermo-optic refractive index change that alters the round-trip phase of the cavity. This results into unwanted feedback dynamics between the optical and mechanical components that can drive the system into strongly nonlinear regimes. Different techniques have been used to mitigate photothermal effects, such as improving the coating quality and reducing thermal stress~\cite{Black04, Evans08}. When these measures are not sufficient, active thermo-compensation becomes a prerequisite to fully stabilise the system~\cite{Hardwick20}. However, this is not always an option for more compact systems~\cite{Kon17}.

Recent research showed the stabilisation of a gram-scale oscillator by the assistance of photothermal feedback~\cite{Altin17}. This technique uses photothermal effects to alter the oscillator's dynamics in a desired manner to passively stabilise the system. A requirement for its success is for the photothermal interaction to cooperate, rather than compete, with the radiation pressure force. This condition, however, may not be naturally satisfied by other optomechanical platforms. Ballmer~\cite{Ballmer15} suggested the idea of inverting the sign of the interaction --- that is, regulate an effective photothermal contraction of the cavity to correspond to an expansion instead --- as a practical process to suppress unintended instabilities. A theoretical proposal has shown that it is possible to use materials with negative photothermal coefficients in the coating layers of the cavity mirrors to cancel out photothermal noise~\cite{Rana16}.

In this paper, we demonstrate a passive technique to achieve cancellation of photothermally induced instability. Our optical system is a vertical Fabry-P\'erot cavity designed for scattering-free optical levitation of a macroscopic mirror~\cite{Guc13, Lecam20}. This platform uses the intra-cavity radiation pressure force to push a milligram-scale mirror against gravity, while taking advantage of the optical spring effect to fully confine the mirror in the three spatial dimensions, in principle free of attachments to any supporting mechanical structures. The high-power requirement for levitation of this relatively large target generates instabilities due to concurring photothermal effects~\cite{Ma20}. We achieve cancellation of these instabilities by inserting an optical window within the cavity, thereby modifying the effective photothermal coefficient of the combined system. Using different window materials, we analyse their photothermal parameters in our cavity-enhanced detection scheme. Our results show a promising route to controllable engineering of optomechanical interactions, establishing photothermal back-action feedback as a convenient stabilisation technique in general optomechanical systems.

\section{Theoretical framework}

\begin{figure}[htbp]
\centering\includegraphics[width=0.6\textwidth]{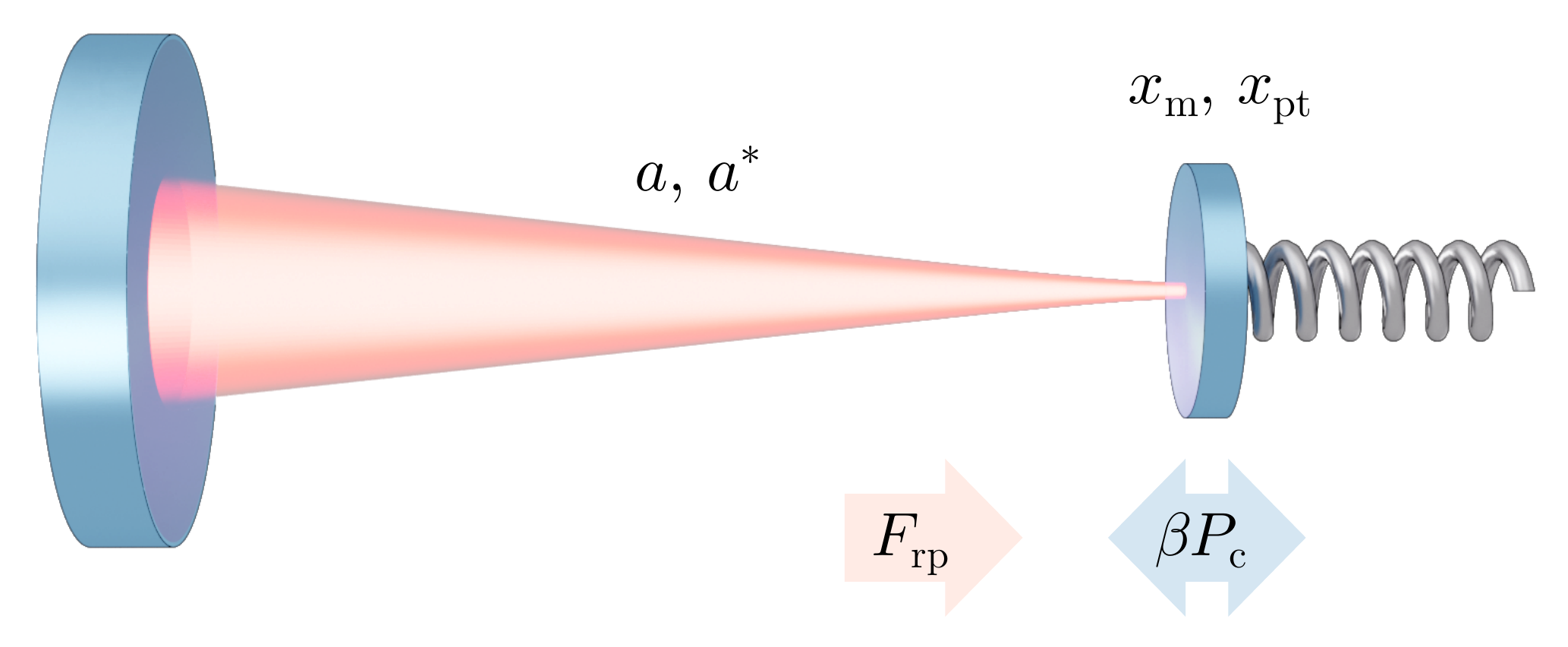}
\caption{Diagram for an optomechanical cavity. The labels highlight the optical field's complex amplitude ($a$, $a^*$) and the oscillator's centre of mass ($x_{\rm m}$) and photothermal ($x_{\rm pt}$) displacements, as well as their interaction through the radiation pressure force ($F_{\rm rp}$) and photothermal effects ($\beta P_{\rm c}$).}
\label{fig: optomechanics}
\end{figure}

We model a linear optical cavity, where the input mirror is fixed and the opposite end mirror is a mechanical oscillator moving on the $x$-axis, as shown in Fig.~\ref{fig: optomechanics}. We consider both radiation pressure (RP) and photothermal (PT) effects to be present. In this configuration, the radiation pressure force pushes outwards, always increasing the cavity length for increasing power. The photothermal effects introduce additional back-action and also modify the effective cavity length. In this model, the direction of change is determined by the sign of the interaction. The full behaviour of the combined system is characterized by the following equations of motion in the rotating frame of the cavity resonant frequency~\cite{Ma20}:

\begin{align}
	\dot{a} & = \phantom{-}[-\kappa+i(\Delta+G(x_{\rm pt}+x_{\rm m})]a+\sqrt{2\kappa_{\rm in}}a_{\rm in}, \label{eq: a}\\
	\dot{a}^{*} & = \phantom{-}[-\kappa-i(\Delta+G(x_{\rm pt}+x_{\rm m})]a^{*}+\sqrt{2\kappa_{\rm in}}a^{*}_{\rm in}, \label{eq: a*}\\
	\dot{x}_{\rm pt} & = -\gamma_{\rm pt}[x_{\rm pt}+\beta(\frac{1}{2}\hbar c G|a|^2)], \label{eq: x_pt}\\
	\dot{x}_{\rm m} & = \phantom{-}p_{\rm m}/m_{\rm eff}, \label{eq: x_m}\\
	\dot{p}_{\rm m} & = -\gamma_{\rm m} p_{\rm m}-m_{\rm eff} \omega_{\rm m}^2 x_{\rm m}+\hbar G|a|^2. \label{eq: p_m}
\end{align}

Here, $a$ is the optical cavity field amplitude, $x_{\rm m}$ is the position of the mechanical oscillator in the cavity, and $x_{\rm pt}$ is the cavity length difference due to the photothermal response. Equations~\ref{eq: a} and~\ref{eq: a*} represent the evolution of the optical field. The cavity detuning is defined as $\Delta = \omega_{\rm o} - \omega_{\rm c}$, where $\omega_{\rm o}$ and $\omega_{\rm c}$ are the angular frequencies of the optical field and cavity resonance, respectively. The parameter $a_{\rm in}$ indicates the amplitude of the external input field, with $\kappa_{\rm in}$ being its input coupling rate. The cavity half-linewidth is $\kappa$ (with $\kappa\gtrsim\kappa_{\rm in}$). The coefficient $G=\omega_{\rm c}/L$ represents the linear optomechanical coupling strength, where $L$ is the length of the cavity at rest. Equation~\ref{eq: x_pt} describes the photothermal effects, following an empirical model for this type of interaction in optomechanical systems~\cite{Kon17, Marino06}. Here, $\beta$ and $\gamma_{\rm pt}$ are, respectively, the susceptivity coefficient and relaxation rate of the photothermal effects. Equations~\ref{eq: x_m} and~\ref{eq: p_m} describe the motion of the mechanical oscillator of effective mass $m_{\rm eff}$, with $\omega_{\rm m}$ being the mechanical angular frequency and $\gamma_{\rm m}$ the mechanical damping rate. The effective mass $m_{\rm eff}$, that is related to the mechanical mode of the mirror, is usually different to the full mass $m$. In this system of equations, we write explicitly the mechanical restoring force $F_{\rm m} = -m_{\rm eff}\omega_{\rm m}^2 x_{\rm m}$, the radiation pressure force $F_{\rm rp} = \hbar G |a|^2$, and the intracavity power $P_{\rm c} = \hbar c G|a|^2/2$ (where $\hbar$ is the reduced Planck constant, $c$ is the speed of light)~\cite{Lecam20, PTIT}.

Some photothermal phenomena, such as thermal expansion of the mirror's coating or substrate due to heat from absorbed optical power, will decrease the cavity length for increasing power, thus competing with radiation pressure. Other phenomena can instead increase the effective cavity length for increasing power --- for example, a positive thermo-optic change in refractive index in response to temperature. Although more than one effect can be present concurrently~\cite{Farsi12}, the phenomenological traits of these interactions are determined by the sign of the effective photothermal coefficient $\beta$. For $\beta>0$, the photothermal effect competes with and opposes the radiation pressure, which can aggravate instabilities in the system. For $\beta<0$, the photothermal effect cooperates with the radiation pressure effect causing an expansion of the cavity length change in the same direction. In this regime, since they contribute differently to the `stiffness' and `damping' properties of the system, they can truly operate together to stabilise the optomechanical cavity from both static and dynamic perspectives. 

\begin{figure}[htbp]
\centering\includegraphics[width=0.8\textwidth]{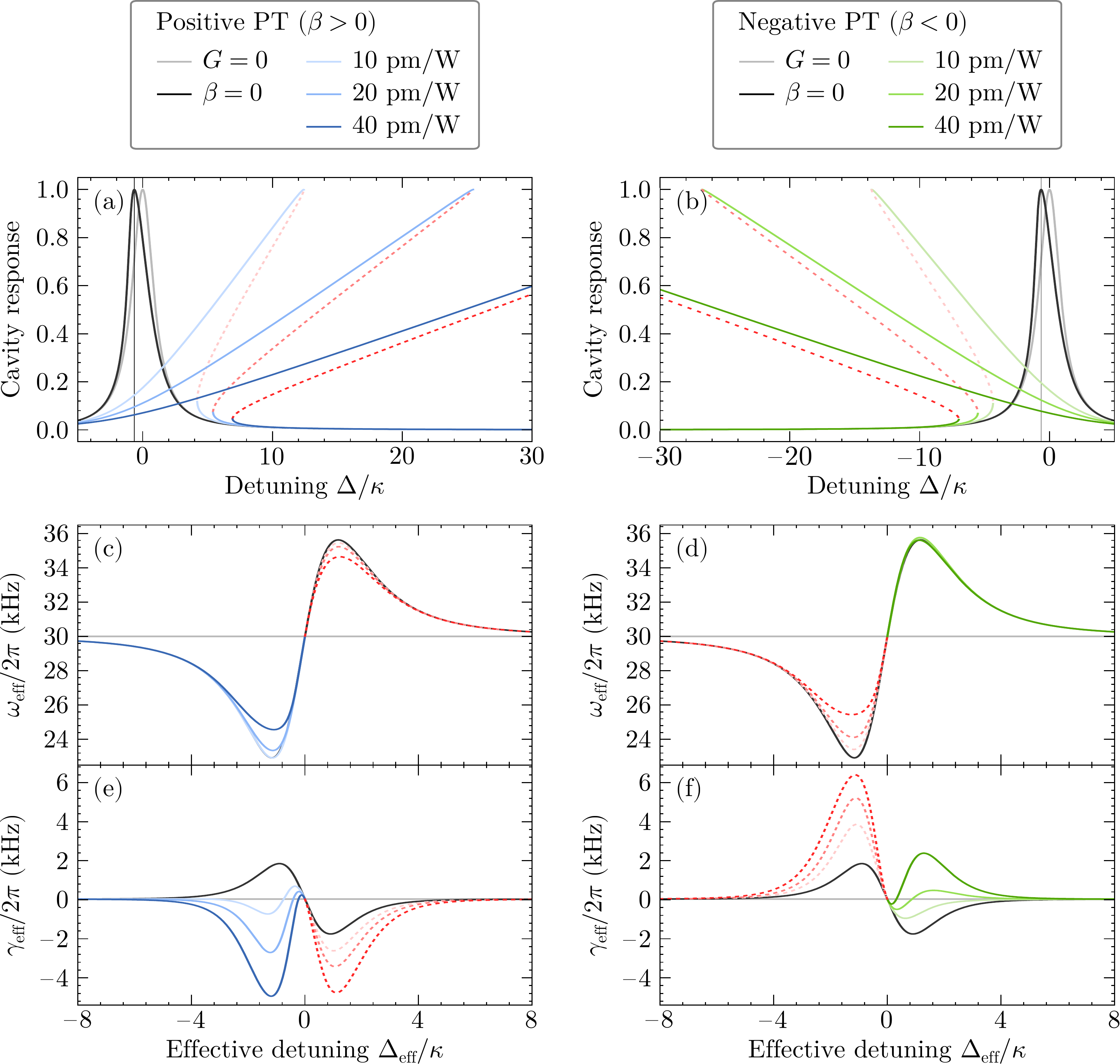} 
\caption{Back-action of radiation pressure and photothermal effects in the optomechanical system. All plots are repeated for progressively stronger photothermal effects corresponding to $|\beta|$ of $10$, $20$, and $40$ \si{\pico\meter\per\watt}, with $\beta>0$ on the left-hand side (solid blue) and $\beta<0$ on the right-hand side (solid green). The black curves represent the same optomechanical system with only radiation pressure and no photothermal effects ($\beta=0$). The reference scenario with no optomechanical back-action at all ($G=0$) is also shown in light grey. \textbf{(a--b)} Normalized cavity response as a function of optical detuning, showing optical bistability as a result of the interaction. Optically unstable solutions are plotted in dashed red --- these are inaccessible steady state solutions, independent of the optomechanical instability resulting from parametric gain ($\gamma_{\rm eff}<0$). The black curve shows that radiation pressure alone also has some back-action, leading to a small shift in resonance (indicated by the solid vertical lines). \textbf{(c--f)} Modification of the oscillator's frequency (subplots (c) and (d)) and damping (subplots (e) and (f)) due to optical back-action. The parameters are plotted as a function of effective detuning, $\Delta_{\rm eff}$. The dashed red sections show effective detunings that are unfeasible because of optical instability. With no back-action (grey curves), $\omega_{\rm eff} = \omega_{\rm m}$ and $\gamma_{\rm eff} = \gamma_{\rm m}$ at all detunings. For all plots, the values of the relevant parameters are: $P_{\rm in} = \SI{200}{\milli\watt}$,  $L = \SI{80}{\milli\metre}$, $\omega_{\rm c}/2\pi = \SI{285.7}{\tera\hertz}$, $\kappa/2\pi = \SI{330}{\kilo\hertz}$, $\kappa_{\rm in}/2\pi = \SI{110}{\kilo\hertz}$, effective mass of the acoustic drum mode as $m_{\rm eff} = \SI{0.38}{\milli\gram}$, $\omega_{\rm m}/2\pi = \SI{30}{\kilo\hertz}$, $\gamma_{\rm m}/2\pi = \SI{30}{\hertz}$, $\gamma_{\rm pt}/2\pi = \SI{400}{\hertz}$.}
\label{fig: back-action}
\end{figure}

In the absence of any optomechanical interaction ($G=0$), the optical intensity of the cavity would follow a Lorentzian response as a function of cavity detuning. However, by letting the optical field interact with the position degrees of freedom, the back-action feedback induces a modified resonance profile as shown in Fig.~\ref{fig: back-action}a--b. When photothermal effects dominate, the cavity resonance is blue-detuned to higher frequencies if $\beta$ is positive, or red-detuned to lower frequencies if $\beta$ is negative. Bistability is observed when the interaction is sufficiently strong~\cite{Agrawal79}, allowing simultaneously one unstable (highlighted in dashed red) and two stable steady-state solutions (solid blue or green) for the same detuning. The cavity response during a detuning scan will therefore depend on the direction of the scan. When $\beta>0$, for example, a scan from red to blue detunings will cause the resonance to follow along and lead the cavity to ``self-lock'' (i.e.\ the change in cavity length occurs in the same direction as the scan). Scanning in the opposite direction, from blue to red detunings, will encounter the unstable region first and the cavity reacts by rapidly crossing over to the opposite side of resonance as an ``anti-locking'' mechanism (i.e.\ the cavity length change opposes the scan, therefore crossing resonance more rapidly). The flipped scenarios will play out if $\beta<0$.

Optical bistability describes only the static steady-state behaviour of the system. The full dynamic evolution has to account for the time-varying elements such as the build-up of oscillations due to the positive restoring force and possible anti-damping from a negative effective viscous coefficient. The Jacobian matrix can be used to analyse the stability of the system. It is obtained from linearisation of Eqs.~\ref{eq: a}--\ref{eq: p_m} as follows:
\begin{align}
M_{\rm J}&=\left[\begin{array}{ccccc}
-\kappa+i\Delta_{\rm eff} & 0 & i G \alpha & i G \alpha & 0\\
0 & -\kappa -i\Delta_{\rm eff}& -i G \alpha^{*} & -i G \alpha^{*} & 0\\
-(c/2)\hbar G\sigma_{\rm pt}\alpha^{*} & -(c/2)\hbar G\sigma_{\rm pt}\alpha &  -\gamma_{\rm pt} & 0& 0\\
0 &0&0&0& 1/m_{\rm eff} \\
\hbar G \alpha^{*}& \hbar G\alpha&0&-m_{\rm eff}\omega_{m}^{2} & -\gamma_{m}\\
\end{array}\right]. \label{eq: jac_m}
\end{align}
Here, $\alpha = \sqrt{2\kappa_{\rm in}} \langle a_{\rm in} \rangle / (\kappa+i\Delta_{\rm eff})$ is the steady-state value of the intra-cavity field amplitude. The effective detuning is $\Delta_{\rm eff} = \Delta+G(\langle x_{\rm pt} \rangle +\langle x_{\rm m} \rangle)$. The angled brackets denote steady-state values for the respective degrees of freedom. The photothermal parameter is defined as $\sigma_{\rm pt} = \gamma_{\rm pt} \beta$. 

The eigenvalues of this matrix contain the dynamical information of the system. The system is considered stable if the real parts of all eigenvalues of the Jacobian matrix are negative. If any eigenvalue has a positive real component, the system is unstable and will not reach a steady-state solution. More specifically, the real parts of the eigenvalues coincide with the damping (or anti-damping) coefficients of the system, and their sign determines whether or not the modes converge to the steady-state solution. The imaginary parts correspond to the steady-state eigenfrequencies in the system. Those eigenvalues can be distinguished to corresponding system parameters by referring to their natural values obtained in the limit of no interaction between optical and mechanical modes ($G = 0$). The optical modification of these mechanical constants corresponds to a generalised \text{optical spring} effect~\cite{Ma21}, in affinity to the similar phenomenon in optomechanical systems where pure radiation pressure force modifies the mechanical susceptibility. Here, both radiation pressure and photothermal forces act together to modify the mechanical response. We therefore have the effective oscillator frequency $\omega_{\rm eff}$ and damping rate $\gamma_{\rm eff}$ as follows:

\begin{align}
\omega_{\rm eff} & = \Im\{\text{eig}(M_{{\rm J}})\}, \label{eq: omega_eff}\\
\gamma_{\rm eff} & = -2\Re\{\text{eig}(M_{{\rm J}})\}, \label{eq: gamma_eff}
\end{align}
\noindent
where $\Im$ is imaginary part, $\Re$ is real part. The effects of the optical interaction on the original frequency and damping rate of the oscillator are shown in Fig.~\ref{fig: back-action}c--f, with $\beta>0$ on the left and $\beta<0$ on the right. We see that the frequency of the oscillator is modified predominantly by radiation pressure, with small change depending on the sign or strength of the photothermal interaction. However, there is a difference in terms of optical stability. For positive $\beta$, the overall system cannot be stable at positive effective detunings. When $\beta$ is negative this blue-detuned regime is instead stable, and the optical back-action gives a positive restoring force (Fig.~\ref{fig: back-action}d) which can be important for free-mass and low-frequency optomechanical systems relying on the optical spring effect for optical confinement. The levitation system investigated in the next section is one such example.

Photothermal interaction has a much stronger influence on the damping properties of the system. Intuitively, this is expected as the photothermal change due to optical power enters directly into the first-order time derivative of displacement (Eq.~\ref{eq: x_pt}). Starting from a natural damping of $\SI{30}{\hertz}$, the effective damping can be orders of magnitude stronger and play a compelling role in inducing parametric instability ($\gamma_{\rm eff}<0$ in Fig.~\ref{fig: back-action}e) or in helping the system reach its steady state ($\gamma_{\rm eff}>0$ in Fig.~\ref{fig: back-action}f) over almost the entire range of effective detuning.

On the whole, both radiation pressure and photothermal forces contribute to optical back-action that modifies the response of the mechanical oscillator within the optomechanical cavity. The two phenomena, however, address different elements of the dynamical evolution of the system. As such, they can in principle contribute in parallel to the stability of the combined optomechanical system. In the presence of more than one photothermal effect --- for example, both material expansion and thermo-optics refractive index change --- it becomes necessary to account for the different time scales by modelling independent equations of motion for the corresponding variables. However, we found that the results are consistent with using a single effective equation in the following conditions: two photothermal effects have similar relaxation rate but different coefficients, or one photothermal effect is dominant. Under these conditions we can consider modifying the cavity with a supplementary photothermal degree of freedom to passively control the effective interaction to be unconditionally stable.

\section{Experimental setup}

\begin{figure}[htbp]
\centering\includegraphics[width=0.9\textwidth]{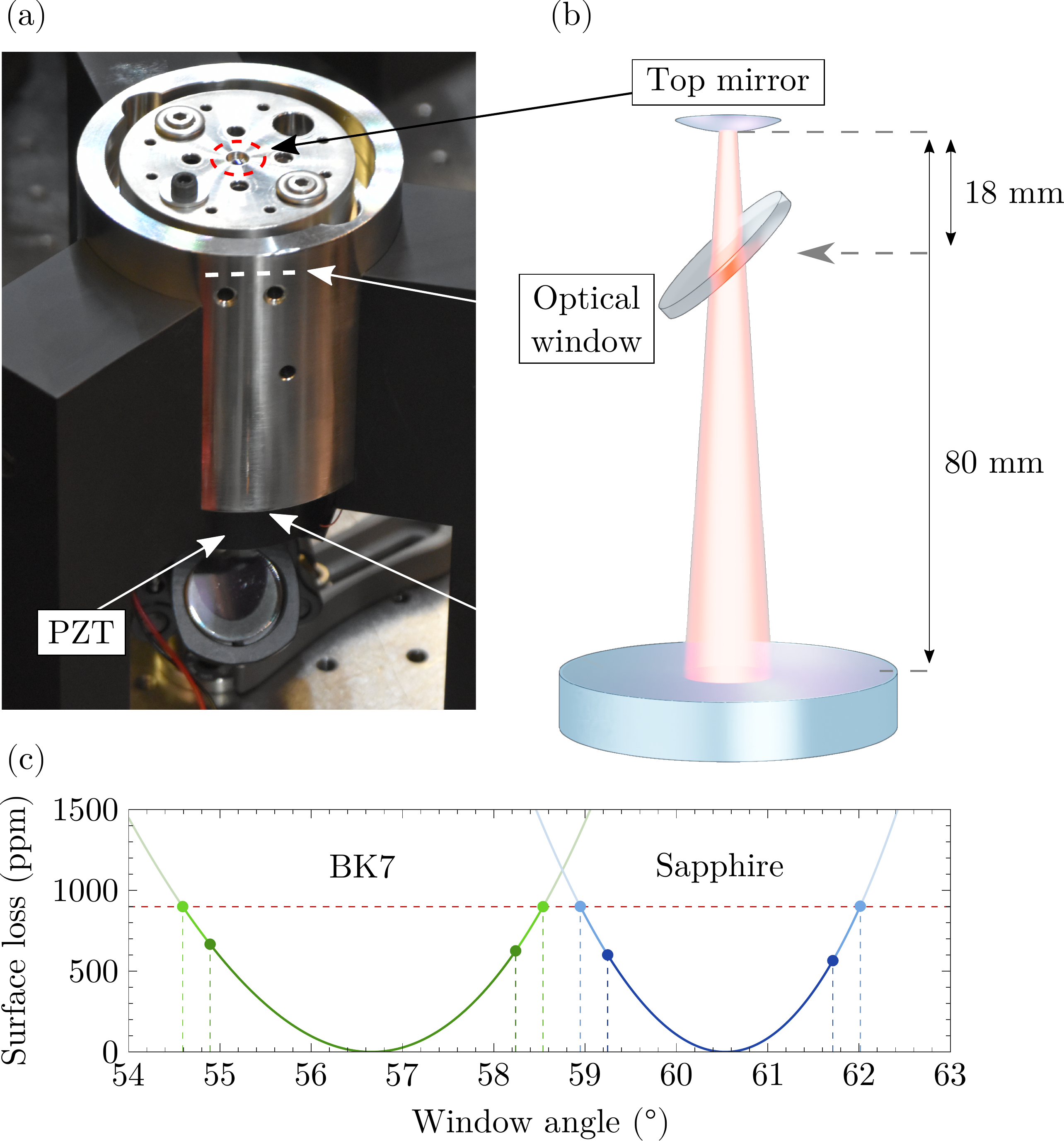} 
\caption{\textbf{(a)} Experimental setup showing the monolithic Invar casing of the optical cavity. \textbf{(b)} Schematic illustration of the vertical optical cavity with the free-standing top mirror and the intra-cavity optical window for photothermal cancellation. The window is tilted at the Brewster's angle to minimise power loss. The beam size, divergence and positions relative to the cavity length are to scale. The beam divergence at the location of the optical window is \SI{0.6}{\degree}. \textbf{(c)} Surface loss at the window's interfaces as a function of the window angle with respect to the laser beam. Brewster's angles are found to be \SI{56.7}{\degree} for N-BK7 and \SI{60.5}{\degree} for sapphire windows. The dashed red line defines a loss threshold to prevent the cavity finesse to lower by more than 30\%. The inner sets of dots mark the acceptable range after taking the beam's divergence into consideration.}
\label{fig: setup}
\end{figure}

Optical levitation is an effective way to decouple the environmental noise and enhance the performance of the mechanical oscillators. A scattering-free optical tripod system for macroscopic levitation has been proposed as a promising platform for quantum metrology and gravitational sensing~\cite{Guc13}. This system was recently trialled with a free-standing mirror acting as the top reflector of a vertical optical cavity~\cite{Ma20}. The underlying radiation pressure from the cavity field provides the necessary force to suspend the mirror against gravity. At the same time, the optical spring effect provides a restoring force to spatially confine the suspended mirror. In practice, however, the smallest amount of absorption in the coating results in photothermal expansion and thermo-optic refractive index change that decrease the effective cavity length and compete with radiation pressure. With around $\SI{3}{\mega\watt\per\centi\metre\squared}$ of optical intensity to suspend the milligram-scale mirror, this effect is critical as it results in the uncontrolled parametric amplification of the mirror's oscillation. In these conditions, the system is unstable and unfit for its designed applications.

In the previous section we discussed how photothermal effects with negative $\beta$ could stabilise the optomechanical cavity by inducing positive damping. To counterbalance the original photothermal effect in the levitation setup, which corresponds to a positive $\beta$, we tested a similar vertical cavity configuration with the addition of an intra-cavity optical window as shown in Fig.~\ref{fig: setup}. The window modifies the overall photothermal response of the system, compensating for the expansion of the mirror with its own and providing an effectively negative photothermal interaction.

Our vertical cavity is a reduced version of optical tripod as a testbed for levitation~\cite{Ma20}. It uses a fused silica spherical cap with a mass of \SI{1.116\pm 0.003}{\milli\gram} as the levitation mirror at the top. This mirror has a diameter of \SI{3}{\milli\metre}, thickness of approximately \SI{50}{\micro\metre}, and radius of curvature of \SI{25}{\milli\metre}. It has an ion beam sputtered coating on the convex surface for high-reflectivity of \SI{99.992}{\percent}. The lower end of the cavity is capped by a conventional high-reflective 1-inch concave mirror (with reflectivity greater than $\SI{99.9}{\percent}$) mounted on a piezoactuator to enable detuning scans. The cavity is \SI{80}{\milli\metre} long and it is enclosed in an Invar bulk spacer to decouple seismic noise. The top mirror is free-standing on a hole that is specifically designed to have three contact points to reduce Van der Waals interactions. This also provides two supporting points at the mirror edge when it is lifted off by the radiation pressure.

Optical loss has to be minimal to achieve sufficient laser power for levitation. In the absence of the laser window, we estimate the total cavity loss to be \SI{2200}{ppm}. This includes the transmissivity of \SI{750}{ppm} for the bottom input mirror, \SI{80}{ppm} for the top levitation mirror, as well as losses of \SI{1370}{ppm} due to optical scattering. The implementation of an intra-cavity optical window introduces additional loss to the optical field, not only by absorption through the material but primarily through scattering at the two surfaces. Although we also experimented with anti-reflection coated windows in a flat horizontal position, best performance was obtained by using the principle of Brewster's angle on uncoated windows. The Invar stage was designed with an adjustable platform to accommodate the optical window at Brewster's angle (as shown in Fig.~\ref{fig: setup}b) to minimise surface scattering loss of $p$-polarised light inside the cavity. In Fig.~\ref{fig: setup}c we calculate the surface loss as a function of tilt angle based on Snell's law and Fresnel equations, showing a reasonable angular range for which the total loss does not exceed \SI{900}{ppm} even after accounting for the beam's divergence angle of ~\SI{0.6}{\degree} (\SI{0.3}{\degree} relative to the vertical axis of the cavity). We tested two different low-absorption materials for our wavelength of \SI{1050}{\nano\metre}: N-BK7 and sapphire. Their refractive indices allowed a reasonable range of more than \SI{2}{\degree} in each case: 54.9--58.2\si{\degree} and 59.2--61.7\si{\degree}, respectively. Residual surface loss proved hard to estimate quantitatively since it depends on the local surface roughness of the window. Internal absorption loss was also minimised by careful choice of the window's material, although we note that some of absorption is necessary in order to modify the effective photothermal interaction. The linear polarisation of the input laser was tuned with a zero-order $\lambda/2$ waveplate to match the point of optimal transmission through the window. Note that the window was positioned closer to the beam's waist to increase optical intensity and therefore obtain higher contribution of photothermal effects for the given absorption loss. Also, since the photothermal rate depends on beam size, this strategy ensured that the relaxation rates of the different photothermal effects on the window and on the top mirror were comparable.

\section{Photothermal cancellation}

\begin{figure}[htbp]
\centering\includegraphics[width=1\textwidth]{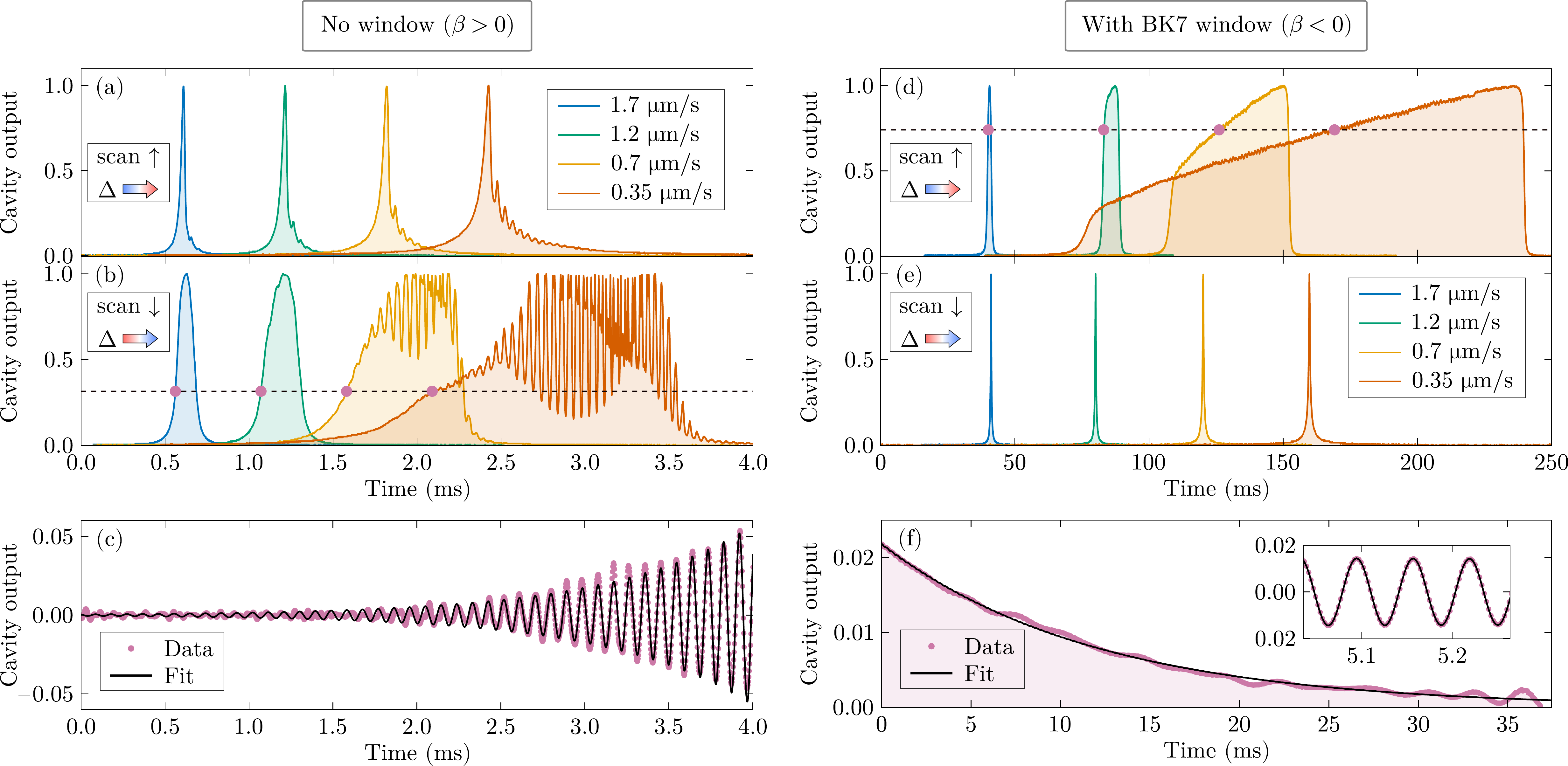} 
\caption{Experimental data showing the modification of cavity dynamics subject to photothermal effects. \textbf{(a--c)} Normalised transmission output of the bare optical cavity without any optical window ($\sigma_{\rm pt}/2\pi=\SI{3750 \pm 650}{\hertz\pico\metre\per\watt}$). The data was taken for \SI{144}{\milli\watt} of input power. The first two panels show optical bistability of the cavity corresponding to: (a) an upward scan of the bottom piezoactuator, from blue to red detunings; (b) a downward scan, from red to blue detunings. The linear sweeps are repeated at different speeds of the piezo-mounted bottom mirror as indicated by the colour legend. Slower scans show evident self-locking when the red-detuned side of resonance is first encountered, with growing oscillations due to parametric instability in this regime. The amplification of the oscillations is shown more clearly in the bottom panel (c), with the cavity fixed at the red-detuned frequency indicated by the purple dots in (b). The purple data is well fitted by a growing exponential oscillation (black curve). \textbf{(d--f)} Normalised transmission output of the same cavity after the inclusion of the 1 mm thick N-BK7 optical window ($\sigma_{\rm pt}/2\pi=\SI{-22380 \pm 670}{\hertz\pico\metre\per\watt}$). An input power of \SI{310}{\milli\watt} was used for this data. The panels showing detuning sweeps have the same order as before, scanning from blue to red detunings at the top (d) and from red to blue detunings at the bottom (e). The presence of the N-BK7 window flips self-locking to the blue-detuned side of resonance, where it would be expected from ordinary radiation pressure interaction. No anti-damping is detected in the system, which is now collectively stable. The data in the last panel (f) has been taken with the cavity fixed at the blue-detuned frequency indicated by the purple dots in (d). It shows how externally excited oscillations decay over time due to the now positive effective damping. For clarity only the points of the exponential envelope are shown (a sample of the full oscillations and their fit is shown in the inset).}
\label{fig: cancellation}
\end{figure}

The main results are shown in Fig.~\ref{fig: cancellation}, where the dynamics of the optomechanical system are shown both for the original system (left-hand side) and when the optical window is incorporated into the cavity (right-hand side).

On its own, the bare optomechanical cavity without any window  is subject to positive photothermal interaction ($\beta>0$). This is evident when performing linear detuning scans by means of the piezoelectric actuator attached to the bottom input mirror. The back-action in the system induces anti-locking for an upward scan (equivalent to a sweep from blue to red detunings, Fig.~\ref{fig: cancellation}a) and self-locking for a downward scan (red to blue detunings, Fig.~\ref{fig: cancellation}b). In both cases, the optomechanical cavity reveals anti-damping on the red-detuned side, with amplified oscillations of the top mirror that destabilise the combined system. These growing oscillations are a consequence of a negative $\gamma_{\rm eff}$, which results in parametric gain that amplifies even the smallest thermal fluctuations. This behaviour is directly observed in Fig.~\ref{fig: cancellation}c, where the cavity is allowed to evolve in time under the steady drive of a red-detuned input field. The resulting dynamics correspond to $\gamma_{\rm eff}/2\pi = -\SI{106\pm 14}{\hertz}$ and $\omega_{\rm eff}/2\pi=\SI{10.63\pm 0.47}{\kilo\hertz}$ (errors are standard deviations over repeated measurements). Fitting the numerical solutions of Eqs.~\ref{eq: a}--\ref{eq: p_m} to the scan data at different speeds, we estimate the photothermal parameter to be $\sigma_{\rm pt}/2\pi=\SI{3750 \pm 650}{\hertz\pico\metre\per\watt}$.

In our experiment, the largest photothermal cancellation was obtained for a \SI{1}{\milli\metre} thick N-BK7 window ($n=1.52$). One of the first observations is that the scan direction for the self-locking and anti-locking processes is reversed (cf.\ Fig.~\ref{fig: cancellation}d--e). This effect, in agreement with the model shown in Fig.~\ref{fig: back-action}b, occurs because the effective photothermal interaction now leads to an \emph{increase} in the optical path length that cooperates with the radiation pressure. Another important observation is the absence of parametric instability on either side of resonance, again in agreement with the trend shown in Fig.~\ref{fig: back-action}f where $\gamma_{\rm eff}$ becomes positive at all detunings. In particular, $\gamma_{\rm eff}$ is now a genuine damping coefficient which will also suppress external excitations (Fig.~\ref{fig: cancellation}f). The fit of the exponential decay returns a value of $\gamma_{\rm eff}/2\pi = \SI{13.6\pm 0.4}{\hertz}$. Both of these effects --- cooperation with radiation pressure and gain of dynamical stability in the system --- are clear signatures of a strongly negative photothermal interaction. By fitting the scanned traces, we estimate a value of $\sigma_{\rm pt}/2\pi=\SI{-22380 \pm 670}{\hertz\pico\metre\per\watt}$.

Indeed, the photothermal cancellation induced by the \SI{1}{\milli\metre} N-BK7 window appears to be far beyond sufficient for achieving stable levitation due to excessive absorption. 
The added loss is on the order of 7000--8000~\si{ppm}, which reduces the cavity finesse by more than a factor of four, down to $650$. In an attempt to find a material with one or even two orders of magnitude lower absorption, we used a \SI{3}{\milli\metre} sapphire window ($n=1.77$). This material showed much better optical properties, introducing only an additional \SI{600}{ppm} of loss to the cavity. This served as evidence that the Brewster's angle method is effective in mitigating optical losses at the surface. However, this sapphire window was insufficient to reverse the sign of system's photothermal effect. 

Thinner N-BK7 wafer windows with a thickness of \SI{0.22}{\milli\metre} were also tested in the cavity to demonstrate that cancellation can be accomplished without a too high sacrifice in terms of loss. The results for these and all other window types are summarised in Table S1 in the Supplementary Information. One of these wafers is sufficient to reduce the effective photothermal coefficient by nearly half with negligible optical loss. Stacking multiple wafers together with the aid of index matching fluids to preserve the optical mode, we observed a progressive improvement to the point where the cavity manifests threshold behaviour with an almost symmetric shape regardless of the direction of the scan ($\sigma_{\rm pt}$ is close to zero, see Fig. S3 in Supplementary Information). In our previous publications, we have discussed levitation assisted by sideband cooling~\cite{Lecam20} or feedback cooling~\cite{Ma20} at this zero-photothermal regime. The photothermal cancellation, however, did not scale linearly with the number of windows, possibly due to the additional photothermal response of the fluid between each wafer. The effective interaction was not strong enough to render $\gamma_{\rm eff}$ positive and the suppression of parametric gain was insufficient to fully stabilise the system. Notably, the added loss is sufficiently low as 900 ppm for a 4-layer N-BK7 stack.

In summary, the best performance in photothermal cancellation was observed with N-BK7 windows --- both in terms of total cancellation (with a 1 mm window) and cancellation per unit thickness (with a 0.22 mm thin wafer). As the Brewster's angle method can effectively reduce surface optical loss, the desired photothermal parameters and low optical loss could in principle be achieved by use of right substrate materials with suitable thickness. In the next section, we further investigate system dynamics of a macroscopic optical levitation setup to demonstrate the effectiveness of photothermal cancellation in stabilising the system.

\section{Stable optical levitation}

With the proof that photothermal cancellation of instabilities is possible, we apply the concept to a cavity levitation system simulated under ideal conditions to show how it can be free from back-action vulnerabilities and suitable for long-term measurements or other specialised applications. With the implementation of optical dilution and quantum readout, a stable optical levitation platform could be used for quantum-limited sensing and metrology of physical constants~\cite{Gonz21, Michi20}, as a prototype for studies of quantum gravity~\cite{Gro16, Gan16}, or as an adaptable testbed for macroscopic quantum mechanics~\cite{Vanner13, Novotny21, Ho19}.

A cavity for optical levitation without photothermal effects should be self-locking during an ascending scan of the bottom mirror. The idea is that, as the cavity approaches resonance from far-blue detunings, the cavity field builds up and its radiation pressure induces a vertical upward force on the free-standing mirror at the top. With sufficient power this mirror is effectively `picked up' by the optical field, dragging the resonance condition along with the scan as it gets lifted. Although the self-locking traces of the stabilised cavity in Fig.~\ref{fig: cancellation}d would suggest precisely this type of behaviour, we know that in this case the response shown is not a consequence of radiation pressure. The losses introduced by the \SI{1}{\milli\metre} N-BK7 window reduce the cavity finesse too much for the resonant optical field to produce the radiation pressure necessary to lift the mirror. The threshold input power $P_{\rm in}$ required to move the mirror is estimated to be \SI{52}{\watt}, while the current laser source is up to 20 W of output power. 
Under these considerations, we know that the path length is increased due to photothermal effects and not because of optical levitation by radiation pressure. 

\begin{figure}[h!]
\centering\includegraphics[width=1\textwidth]{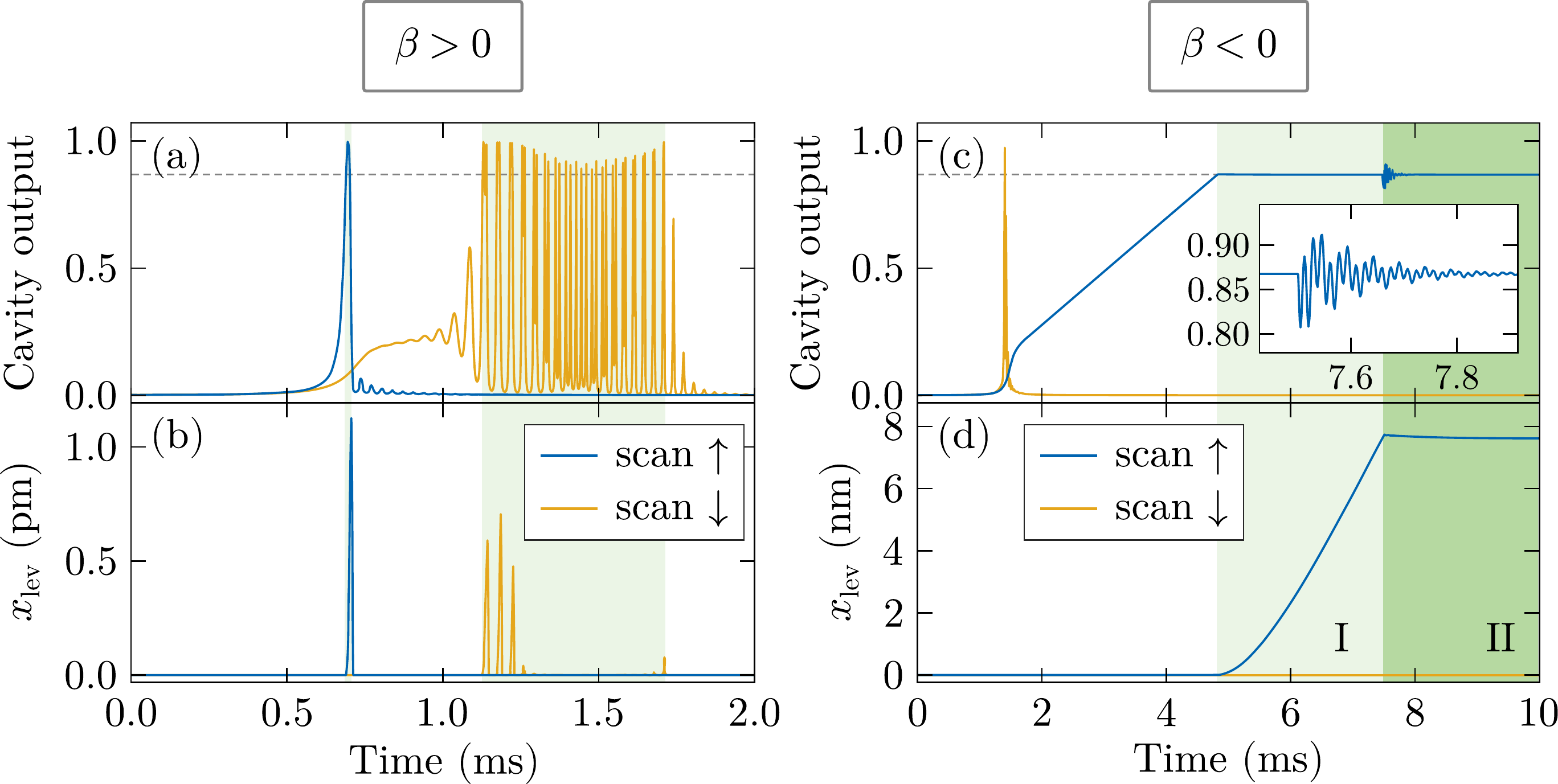}
\caption{Numerical simulation for a levitation system with photothermal effects, for a positive (a--b) or negative (c--d) coefficient $\beta$. (a--b) $\sigma_{\rm pt}/2\pi=\SI{3000}{\hertz\pico\metre\per\watt}$. (c--d) $\sigma_{\rm pt}/2\pi=\SI{-3000}{\hertz\pico\metre\per\watt}$. The top panels show the cavity outputs, which are normalised to the resonance transmissions. The panels at the bottom show the free-standing mirror's centre-of-mass degree of freedom, $x_{\rm lev}$. The simulations are obtained in presence of a linear detuning scan by the input mirror at the bottom, with direction as indicated by the legend. In all panels, the lighter green shaded area indicates where the cavity power exceeds the threshold for levitation. On the right-hand side, the differently shaded areas indicate: I) lifting of the mirror during the scan; II) sustaining levitation of the mirror at the height when the scan is stopped at \SI{7.5}{\milli\second}.}
\label{fig: simulation}
\end{figure}

We consider now the scenario of a cavity with photothermal interaction but also enough optical power to allow for levitation --- either through a reduction in absorption losses, availability of higher input laser power, or the use of an even lighter mirror. To account for the centre-of-mass degree of freedom of the levitated mirror ($x_{\rm lev}$), we include the following equations of motion into the set of Eqs.~\ref{eq: a}--\ref{eq: p_m}:
\begin{align}
	\dot{x}_{\rm lev} & = p_{\rm lev}/m, \label{eq: x_lev}\\
	\dot{p}_{\rm lev} & =
	\begin{cases}	
		-\gamma_{\rm lev}p_{\rm lev} & F_{\rm rp}\leq F_{\rm g}\\
		-\gamma_{\rm lev}p_{\rm lev}+\hbar G|a|^2-mg & F_{\rm rp}>F_{\rm g}
	\end{cases}. \label{eq: p_lev}
\end{align}
The effective detuning for the cavity field $a$ is now considered to be $\Delta_{\rm eff} = \Delta+G(\langle x_{\rm pt} \rangle +\langle x_{\rm m} \rangle + \langle x_{\rm lev} \rangle)$.
The weight of the mirror, $F_{\rm g} = mg$, defines a threshold for the radiation pressure force $F_{\rm rp}$. Here, $m$ is the full mass of the levitated mirror. If the radiation pressure force is not stronger than the gravitational weight, the mirror remains on its stage balanced by the support reaction. When the threshold is surpassed ($\hbar G|a|^2>mg$) the mirror is lifted to a new equilibrium position above the stage, subject to the balance between the two opposing forces. 

The dynamics of the cavity and the free-standing mirror are simulated in Fig.~\ref{fig: simulation}. First we discuss the previous scenario of an empty cavity with positive photothermal interaction ($\sigma_{\rm pt}/2\pi=\SI{3000}{\hertz\pico\metre\per\watt}$), which causes the excitation of acoustic modes of the mirror by parametric gain. We scan the cavity detuning $\Delta$ linearly with time by means of the bottom mirror and observe that, as expected, the photothermal effects modify the optical response so that the system cannot generally maintain a steady state (Fig.~\ref{fig: simulation}a). When the power threshold for levitation indicated by the grey dashed line is exceeded, the mirror (Fig.~\ref{fig: simulation}b) is seen to respond to strong but brief impulse responses, regardless of the scan direction.

The system dynamics are very different when the photothermal interaction is negative ($\sigma_{\rm pt}/2\pi=-\SI{3000}{\hertz\pico\metre\per\watt}$). In this case, self-locking is seen during an upward scan (Fig.~\ref{fig: simulation}c, in accordance with the results presented in Fig.~\ref{fig: cancellation}, but $\sigma_{\rm pt}$ here is nearly 10 times less negative than the \SI{1}{\milli\metre} BK7 window). No  amplification of oscillations due to parametric gain appears in the system. When the cavity reaches the power threshold for levitation, we can see that it automatically locks to this off-resonant detuning. Correspondingly, in Fig.~\ref{fig: simulation}d we see that the mirror `latches' to the scan and starts ascending (first shaded region, lighter green), compensating for the detuning change imposed by the scan. To demonstrate self-supporting stability, we stop the scan at the \SI{7.5}{\milli\second} (second shaded region, darker green). After registering some minor perturbations (inset), the cavity stablises on its own to the same steady-state value and continues to provide the required radiation pressure force for levitation. The mirror, balanced by the gravitational and the optical force, stops at the height reached (lifted by $\sim$\SI{8}{\nano\meter}) when the scan is stopped and stably levitates over time. An example of levitation through a smaller photothermal cancellation ($\sigma_{\rm pt}/2\pi=-\SI{1000}{\hertz\pico\metre\per\watt}$) is also given in Supplementary Information (Fig.~S4). This theoretical analysis shows that an optical window with negative photothermal coefficient can provide the necessary damping to suppress undesired excitations and stabilise the optical spring designed to trap the levitated mirror. If we successfully achieve full tripod levitation with this photothermal technique, we should consider Brownian noise as well as other noise sources in the future.

\section{Conclusions}

In this paper, we considered the impact of photothermal effects in optomechanical cavities and their implications with regards to the underlying stability of the system. By modifying the photothermal response of a cavity, we illustrated the principle of photothermal cancellation whereby the dynamics of a formerly unstable system are rendered steady by the suppression of parametric gain. The result is a reciprocal cooperation between radiation pressure and photothermal interaction, fit for high-sensitivity resonating or compact systems in many areas of optics, optomechanics, photonics and laser technologies.

The technique demonstrated in this paper accomplishes photothermal cancellation by introducing an additional channel of interaction within the system: the photothermal degree of freedom of an optical window internal to the cavity. The manifestation of photothermal cancellation provides proof that passive stabilisation techniques are viable even in the most extreme high-power environments. This provides what we believe, to the best of our knowledge, a new control tool for a variety of optical systems and high-precision metrological applications. Also, a theoretical proposal~\cite{Pinard08} shows that the photothermal effect can bring an oscillator to its quantum ground state in the bad-cavity regime --- when the mechanical resonant frequency is smaller than the cavity decay rate. An extension of this technique for full quantum operations is possible by aiming at the direct cancellation of the inherent photothermal effects in the system instead of a balance between independent effects.

\section*{Funding Information}
This research was funded by Centre of Excellence for Quantum Computation and Communication Technology, Australian Research Council (CE170100012); Australian Government Research Training Program Scholarship; Australian Research Council Laureate Fellowship (FL150100019).

%
%

\section*{Disclosures}
%
%
%
%

\medskip

The authors declare no conflicts of interest related to this article.

\section*{Data availability}
Data underlying the results presented in this paper are not publicly available at this time but may be obtained from the authors upon reasonable request.

\section*{Supplemental Documents}
See Supplementary Information for supporting content.




\begin{thebibliography}{}
\newcommand{\enquote}[1]{``#1''}

\end{thebibliography}


\begin{thebibliography}{10}
\newcommand{\enquote}[1]{``#1''}

\bibitem{Abb16}
B.~P. Abbott, R.~Abbott, T.~Abbott, M.~Abernathy, F.~Acernese, K.~Ackley,
  C.~Adams, T.~Adams, P.~Addesso, R.~Adhikari \emph{et~al.}, \enquote{GW150914:
  The advanced LIGO detectors in the era of first discoveries,} Physical Review
  Letters \textbf{116}, 131103 (2016).

\bibitem{Min20}
M.~Ming, Y.~Luo, Y.-R. Liang, J.-Y. Zhang, H.-Z. Duan, H.~Yan, Y.-Z. Jiang,
  L.-F. Lu, Q.~Xiao, Z.~Zhou \emph{et~al.}, \enquote{Ultraprecision
  intersatellite laser interferometry,} International Journal of Extreme
  Manufacturing \textbf{2}, 022003 (2020).

\bibitem{Teu11}
J.~D. Teufel, T.~Donner, D.~Li, J.~W. Harlow, M.~Allman, K.~Cicak, A.~J.
  Sirois, J.~D. Whittaker, K.~W. Lehnert, and R.~W. Simmonds, \enquote{Sideband
  cooling of micromechanical motion to the quantum ground state,} Nature
  \textbf{475}, 359--363 (2011).

\bibitem{Sch17}
R.~Schnabel, \enquote{Squeezed states of light and their applications in laser
  interferometers,} Physics Reports \textbf{684}, 1--51 (2017).

\bibitem{Evg19}
E.~V. Mikheev, A.~S. Pugin, D.~A. Kuts, S.~A. Podoshvedov, and N.~B. An,
  \enquote{Efficient production of large-size optical schr{\"o}dinger cat
  states,} Scientific Reports \textbf{9}, 1--15 (2019).

\bibitem{Guc20}
G.~Guccione, T.~Darras, H.~Le~Jeannic, V.~B. Verma, S.~W. Nam,
  A.~Cavaill{\`e}s, and J.~Laurat, \enquote{Connecting heterogeneous quantum
  networks by hybrid entanglement swapping,} Science Advances \textbf{6},
  eaba4508 (2020).

\bibitem{Asp14}
M.~Aspelmeyer, T.~J. Kippenberg, and F.~Marquardt, \enquote{Cavity
  optomechanics,} Reviews of Modern Physics \textbf{86}, 1391 (2014).

\bibitem{Liu12}
Y.~Liu, H.~Miao, V.~Aksyuk, and K.~Srinivasan, \enquote{Wide cantilever
  stiffness range cavity optomechanical sensors for atomic force microscopy,}
  Optics Express \textbf{20}, 18268--18280 (2012).

\bibitem{Lec15}
F.~Lecocq, J.~B. Clark, R.~W. Simmonds, J.~Aumentado, and J.~D. Teufel,
  \enquote{Quantum nondemolition measurement of a nonclassical state of a
  massive object,} Physical Review X \textbf{5}, 041037 (2015).

\bibitem{Chan11}
J.~Chan, T.~Alegre, A.~H. Safavi-Naeini, J.~T. Hill, A.~Krause,
  S.~Gr{\"o}blacher, M.~Aspelmeyer, and O.~Painter, \enquote{Laser cooling of a
  nanomechanical oscillator into its quantum ground state,} Nature
  \textbf{478}, 89--92 (2011).

\bibitem{Ashkin87}
A.~Ashkin and J.~M. Dziedzic, \enquote{Optical trapping and manipulation of
  viruses and bacteria,} Science \textbf{235}, 1517--1520 (1987).

\bibitem{Anderegg19}
L.~Anderegg, L.~W. Cheuk, Y.~Bao, S.~Burchesky, W.~Ketterle, K.-K. Ni, and
  J.~M. Doyle, \enquote{An optical tweezer array of ultracold molecules,}
  Science \textbf{365}, 1156--1158 (2019).

\bibitem{Garces02}
V.~Garc{\'e}s-Ch{\'a}vez, D.~McGloin, H.~Melville, W.~Sibbett, and K.~Dholakia,
  \enquote{Simultaneous micromanipulation in multiple planes using a
  self-reconstructing light beam,} Nature \textbf{419}, 145--147 (2002).

\bibitem{Li11}
T.~Li, S.~Kheifets, and M.~G. Raizen, \enquote{{Millikelvin cooling of an
  optically trapped microsphere in vacuum},} Nature Physics \textbf{7},
  527--530 (2011).

\bibitem{Del20}
U.~Deli{\'c}, M.~Reisenbauer, K.~Dare, D.~Grass, V.~Vuleti{\'c}, N.~Kiesel, and
  M.~Aspelmeyer, \enquote{Cooling of a levitated nanoparticle to the motional
  quantum ground state,} Science \textbf{367}, 892--895 (2020).

\bibitem{Corbitt07}
T.~Corbitt, Y.~Chen, E.~Innerhofer, H.~M{\"u}ller-Ebhardt, D.~Ottaway,
  H.~Rehbein, D.~Sigg, S.~Whitcomb, C.~Wipf, and N.~Mavalvala, \enquote{An
  all-optical trap for a gram-scale mirror,} Physical Review Letters
  \textbf{98}, 150802 (2007).

\bibitem{Altin17}
P.~Altin, T.-H. Nguyen, B.~Slagmolen, R.~Ward, D.~Shaddock, and D.~McClelland,
  \enquote{A robust single-beam optical trap for a gram-scale mechanical
  oscillator,} Scientific Reports \textbf{7}, 1--8 (2017).

\bibitem{Corbitt2007}
T.~Corbitt, C.~Wipf, T.~Bodiya, D.~Ottaway, D.~Sigg, N.~Smith, S.~Whitcomb, and
  N.~Mavalvala, \enquote{Optical dilution and feedback cooling of a gram-scale
  oscillator to 6.9 mk,} Physical Review Letters \textbf{99}, 160801 (2007).

\bibitem{Guc13}
G.~Guccione, M.~Hosseini, S.~Adlong, M.~Johnsson, J.~Hope, B.~Buchler, and
  P.~K. Lam, \enquote{Scattering-free optical levitation of a cavity mirror,}
  Physical Review Letters \textbf{111}, 183001 (2013).

\bibitem{Singh10}
S.~Singh, G.~Phelps, D.~Goldbaum, E.~Wright, and P.~Meystre,
  \enquote{All-optical optomechanics: an optical spring mirror,} Physical
  review letters \textbf{105}, 213602 (2010).

\bibitem{Zhao15}
C.~Zhao, L.~Ju, Q.~Fang, C.~Blair, J.~Qin, D.~Blair, J.~Degallaix, and
  H.~Yamamoto, \enquote{Parametric instability in long optical cavities and
  suppression by dynamic transverse mode frequency modulation,} Physical Review
  D \textbf{91}, 092001 (2015).

\bibitem{Rok05}
H.~Rokhsari, T.~J. Kippenberg, T.~Carmon, and K.~J. Vahala,
  \enquote{Radiation-pressure-driven micro-mechanical oscillator,} Optics
  Express \textbf{13}, 5293--5301 (2005).

\bibitem{aggarwal2022searching}
N.~Aggarwal, G.~P. Winstone, M.~Teo, M.~Baryakhtar, S.~L. Larson, V.~Kalogera,
  and A.~A. Geraci, \enquote{Searching for new physics with a
  levitated-sensor-based gravitational-wave detector,} Physical Review Letters
  \textbf{128}, 111101 (2022).

\bibitem{Ma20}
J.~Ma, J.~Qin, G.~T. Campbell, G.~Guccione, R.~Lecamwasam, B.~C. Buchler, and
  P.~K. Lam, \enquote{Observation of nonlinear dynamics in an optical
  levitation system,} Communications Physics \textbf{3}, 1--10 (2020).

\bibitem{Evans08}
M.~Evans, S.~Ballmer, M.~Fejer, P.~Fritschel, G.~Harry, and G.~Ogin,
  \enquote{Thermo-optic noise in coated mirrors for high-precision optical
  measurements,} Physical Review D \textbf{78}, 102003 (2008).

\bibitem{Cerdonio01}
M.~Cerdonio, L.~Conti, A.~Heidmann, and M.~Pinard, \enquote{Thermoelastic
  effects at low temperatures and quantum limits in displacement measurements,}
  Physical Review D \textbf{63}, 082003 (2001).

\bibitem{Rosa02}
M.~De~Rosa, L.~Conti, M.~Cerdonio, M.~Pinard, and F.~Marin,
  \enquote{Experimental measurement of the dynamic photothermal effect in
  fabry-perot cavities for gravitational wave detectors,} Physical Review
  Letters \textbf{89}, 237402 (2002).

\bibitem{Evan15}
M.~Evans, S.~Gras, P.~Fritschel, J.~Miller, L.~Barsotti, D.~Martynov,
  A.~Brooks, D.~Coyne, R.~Abbott, R.~X. Adhikari \emph{et~al.},
  \enquote{Observation of parametric instability in advanced ligo,} Physical
  Review lLetters \textbf{114}, 161102 (2015).

\bibitem{stubenvoll2013photothermal}
M.~Stubenvoll, B.~Sch{\"a}fer, K.~Mann, A.~Walter, and L.~Zittel,
  \enquote{Photothermal absorption measurements for improved thermal stability
  of high-power laser optics,} in \enquote{Laser-Induced Damage in Optical
  Materials: 2013,} , vol. 8885 (SPIE, 2013), vol. 8885, pp. 235--245.

\bibitem{jiang2020optothermal}
X.~Jiang and L.~Yang, \enquote{Optothermal dynamics in whispering-gallery
  microresonators,} Light: Science \& Applications \textbf{9}, 1--15 (2020).

\bibitem{qiu2022dissipative}
L.~Qiu, G.~Huang, I.~Shomroni, J.~Pan, P.~Seidler, and T.~J. Kippenberg,
  \enquote{Dissipative quantum feedback in measurements using a parametrically
  coupled microcavity,} PRX Quantum \textbf{3}, 020309 (2022).

\bibitem{Black04}
E.~D. Black, I.~S. Grudinin, S.~R. Rao, and K.~G. Libbrecht, \enquote{Enhanced
  photothermal displacement spectroscopy for thin-film characterization using a
  fabry-perot resonator,} Journal of applied physics \textbf{95}, 7655--7659
  (2004).

\bibitem{Hardwick20}
T.~Hardwick, V.~J. Hamedan, C.~Blair, A.~C. Green, and D.~Vander-Hyde,
  \enquote{Demonstration of dynamic thermal compensation for parametric
  instability suppression in advanced LIGO,} Classical and Quantum Gravity
  \textbf{37}, 205021 (2020).

\bibitem{Kon17}
K.~Konthasinghe, J.~G. Velez, A.~J. Hopkins, M.~Peiris, L.~T. Profeta,
  Y.~Nieves, and A.~Muller, \enquote{Self-sustained photothermal oscillations
  in high-finesse fabry-perot microcavities,} Physical Review A \textbf{95},
  013826 (2017).

\bibitem{Ballmer15}
S.~W. Ballmer, \enquote{Photothermal transfer function of dielectric mirrors
  for precision measurements,} Physical Review D \textbf{91}, 023010 (2015).

\bibitem{Rana16}
T.~Chalermsongsak, E.~D. Hall, G.~D. Cole, D.~Follman, F.~Seifert, K.~Arai,
  E.~K. Gustafson, J.~R. Smith, M.~Aspelmeyer, and R.~X. Adhikari,
  \enquote{Coherent cancellation of photothermal noise in gaas/al0. 92ga0. 08as
  bragg mirrors,} Metrologia \textbf{53}, 860 (2016).

\bibitem{Lecam20}
R.~Lecamwasam, A.~Graham, J.~Ma, K.~Sripathy, G.~Guccione, J.~Qin, G.~Campbell,
  B.~Buchler, J.~J. Hope, and P.~K. Lam, \enquote{Dynamics and stability of an
  optically levitated mirror,} Physical Review A \textbf{101}, 053857 (2020).

\bibitem{Marino06}
F.~Marino, M.~De~Rosa, and F.~Marin, \enquote{Canard orbits in fabry-perot
  cavities induced by radiation pressure and photothermal effects,} Physical
  Review E \textbf{73}, 026217 (2006).

\bibitem{PTIT}
J.~Ma, J.~Qin, G.~T. Campbell, R.~Lecamwasam, K.~Sripathy, J.~Hope, B.~C.
  Buchler, and P.~K. Lam, \enquote{Photothermally induced transparency,}
  Science Advances \textbf{6}, eaax8256 (2020).

\bibitem{Farsi12}
A.~Farsi, M.~Siciliani~de Cumis, F.~Marino, and F.~Marin, \enquote{Photothermal
  and thermo-refractive effects in high reflectivity mirrors at room and
  cryogenic temperature,} Journal of Applied Physics \textbf{111}, 043101
  (2012).

\bibitem{Agrawal79}
G.~Agrawal and H.~Carmichael, \enquote{Optical bistability through nonlinear
  dispersion and absorption,} Physical Review A \textbf{19}, 2074 (1979).

\bibitem{Ma21}
J.~Ma, G.~Guccione, R.~Lecamwasam, J.~Qin, G.~T. Campbell, B.~C. Buchler, and
  P.~K. Lam, \enquote{Optical back-action on the photothermal relaxation rate,}
  Optica \textbf{8}, 177--183 (2021).

\bibitem{Gonz21}
C.~Gonzalez-Ballestero, M.~Aspelmeyer, L.~Novotny, R.~Quidant, and
  O.~Romero-Isart, \enquote{Levitodynamics: Levitation and control of
  microscopic objects in vacuum,} Science \textbf{374}, eabg3027 (2021).

\bibitem{Michi20}
Y.~Michimura and K.~Komori, \enquote{Quantum sensing with milligram scale
  optomechanical systems,} The European Physical Journal D \textbf{74}, 1--14
  (2020).

\bibitem{Gro16}
A.~Gro{\ss}ardt, J.~Bateman, H.~Ulbricht, and A.~Bassi, \enquote{Optomechanical
  test of the schr{\"o}dinger-newton equation,} Physical Review D \textbf{93},
  096003 (2016).

\bibitem{Gan16}
C.~Gan, C.~Savage, and S.~Scully, \enquote{Optomechanical tests of a
  schr{\"o}dinger-newton equation for gravitational quantum mechanics,}
  Physical Review D \textbf{93}, 124049 (2016).

\bibitem{Vanner13}
M.~Vanner, J.~Hofer, G.~Cole, and M.~Aspelmeyer,
  \enquote{Cooling-by-measurement and mechanical state tomography via pulsed
  optomechanics,} Nature communications \textbf{4}, 1--8 (2013).

\bibitem{Novotny21}
F.~Tebbenjohanns, M.~L. Mattana, M.~Rossi, M.~Frimmer, and L.~Novotny,
  \enquote{Quantum control of a nanoparticle optically levitated in cryogenic
  free space,} Nature \textbf{595}, 378--382 (2021).

\bibitem{Ho19}
C.~T.~M. Ho, R.~B. Mann, and T.~C. Ralph, \enquote{Quantum optical levitation
  of a mirror,}  arXiv 1911.02705 (2019).

\bibitem{Pinard08}
M.~Pinard and A.~Dantan, \enquote{Quantum limits of photothermal and radiation
  pressure cooling of a movable mirror,} New Journal of Physics \textbf{10},
  095012 (2008).

\end{thebibliography}

\clearpage
\section*{Supplemental Information}
This document provides supplementary information to ``Cancellation of photothermally induced instability''. Section~1 lists the characterized photothermal parameters for the optical windows tested in our experimental system. Section~2 explains how theoretical fits and error analyses are conducted in different experimental settings. It also shows representative plots for experimental data and theoretical simulation of cavity response subject to photothermal effects. Section~3 gives an example of stable levitation achieved through a small negative photothermal parameter.

\subsection*{1. Optical Window Parameters}
\label{sec: Optical Window Parameters}
Table~\ref{tab: parameters} lists the parameters for the bare cavity and optical windows tested in our experimental system. The \SI{1}{\milli\meter} thick N-BK7 window is produced by Thorlabs (part number WG11010). The sapphire and thin N-BK7 wafer windows are from Edmund Optics (part number 66-188). The N-BK7 windows from different suppliers can have very different optical absorption and photothermal properties. According to transmittance datasheets provided by the suppliers, there is a large difference in the transmittances that can infer very different optical absorptions. The transmittance reported by Thorlabs is 0.927 (10 mm-thick sample) at the wavelength of 1050nm. The transmittance reported by Edmund Optics is 0.999 (10 mm-thick sample) and 0.997 (25 mm-thick sample) at the same wavelength. The N-BK7 $\times$2 and N-BK7 $\times$4 are stacks of individual thin wafers joined by small amounts of optical index-matching fluid for N-BK7 (Cargille's BK7 glass matching liquid, part number 19586). 

\begin{table}[h!]
\centering
\caption{Parameters of the cavity with different optical windows.}
\begin{tabular}{|l|l|l|l|l|l|l|}
\hline
 & Bare & Sapphire & N-BK7$\times$1 & N-BK7$\times$2 & N-BK7$\times$4 & N-BK7 thick \\ \hline
Refractive index& --- & 1.77 & 1.52 & 1.52&1.52 &1.52  \\ \hline
Thickness (\SI{}{\milli\meter}) & --- & 3.0 & 0.22 & 0.44& 0.88 & 1.0 \\ \hline
Photothermal coefficient,&&&&&&\\
$\sigma_{\rm pt}/2\pi$ (\SI{}{\hertz\pico\metre\per\watt})& \SI{3750 \pm 650}{} & \SI{2870 \pm 350}{} &\SI{2120 \pm 740}{}&\SI{ 960\pm 230}{}& N\slash A &  \SI{-22380 \pm 670}{}  \\ \hline
Photothermal susceptivity,&&&&&&\\
$\beta$ (\SI{}{\pico\metre\per\watt})& \SI{9.3 \pm 1.9}{} & \SI{10.5 \pm 1.3}{}  &\SI{7.0\pm 0.6}{}&\SI{6.6\pm 1.5}{}&N\slash A & \SI{-5100 \pm 20}{}\\ \hline
Cavity finesse & 2850 & 2240 &2350&2070&2030& 650  \\ \hline
Cavity linewidth, $\kappa$ (\SI{}{\mega\hertz})& 0.33 & 0.42 &0.40&0.45&0.46 & 1.44\\ \hline
\end{tabular}
\label{tab: parameters}
\end{table}

\subsection*{2. Theoretical Fit and Error Analysis}
\label{sec: Theoretical Fit and Error Analysis}
Raw data traces are strongly affected by background noise sources (e.g. acoustic vibrations, laser fluctuation, and seismic noise) and uncertainties introduced by the nonlinearity of the piezo actuator. Therefore, our regular approach to parameter estimation in the different experimental configurations consisted in taking multiple traces and fitting them individually to our model with least-square optimisation. By doing so on a full data set we obtain a statistical distribution of the best fit parameters. The calibration of scan speed and the measurement of cavity finesse, linewidth, and input power were performed independently so that only $\sigma_{\rm pt}$ and $\beta$ were free parameters during the optimization. This method however was not always possible, and some configurations required a different approach as discussed further below.

\begin{figure}[h!] \centering
\includegraphics[width=1\columnwidth]{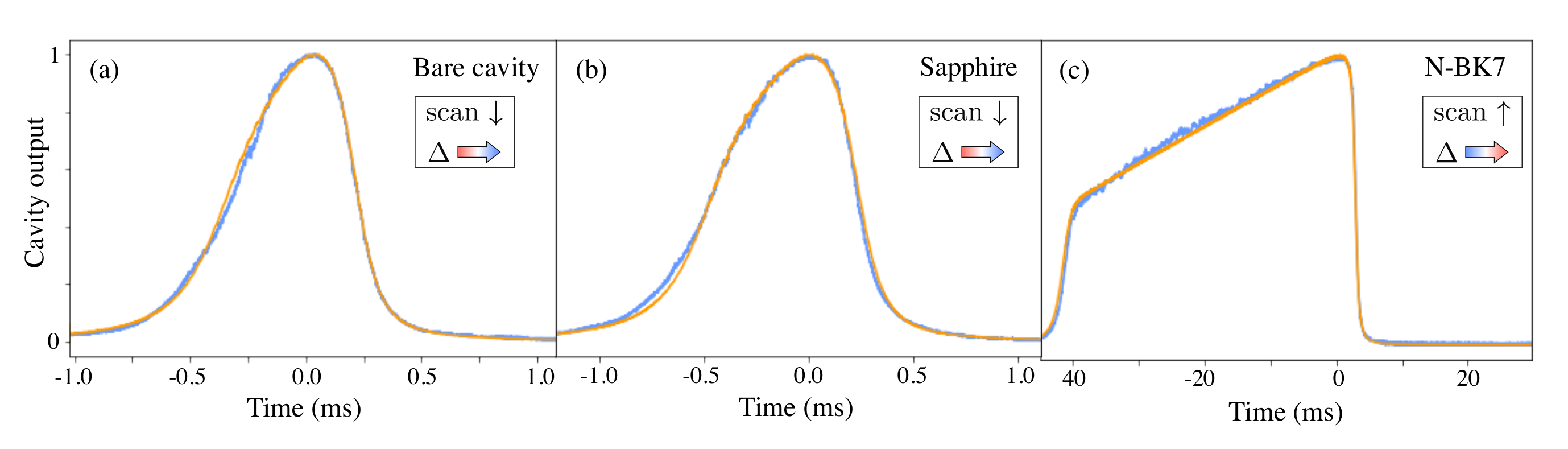} 
\caption{Representative plots of cavity output for (a) the bare cavity, (b) cavity with a \SI{3}{\milli\metre} sapphire window, (c) cavity with \SI{1.0}{\milli\metre} thick N-BK7 window. The blue curves are single experimental traces. The orange curves are best theoretical fits. The resulting parameters are: (a) $\sigma_{\rm pt}/2\pi=\SI{3962}{\hertz\pico\metre\per\watt}$ and $\beta = \SI{9.3}{\pico\metre\per\watt}$; (b) $\sigma_{\rm pt}/2\pi=\SI{2930}{\hertz\pico\metre\per\watt}$ and $\beta = \SI{10.5}{\pico\metre\per\watt}$; (c) $\sigma_{\rm pt}/2\pi=\SI{-21930}{\hertz\pico\metre\per\watt}$ and $\beta = \SI{-5100}{\pico\metre\per\watt}$.}
\label{fig: SI_1}
\end{figure}

For the bare cavity, the cavity with the sapphire window, and the cavity with the \SI{1}{\milli\metre} thick N-BK7 window, we mapped the best fits for $\sigma_{\rm pt}$ and $\beta$ from the raw single traces. The numbers reported in the main paper and in Table~\ref{tab: parameters} are obtained from the average best fit results, with error given by the standard deviation. For these configurations only the self-locking time series for a downward scan of the bottom mirror were considered as they allowed better precision in identifying the appropriate regime. Because the photothermal parameter for \SI{1}{\milli\metre} N-BK7 window is negative, its scan direction is upwards for showing the self-locking side. Figure~\ref{fig: SI_1} shows representative plots corresponding to single data traces and their optimal theoretical fits. 

With thin N-BK7 and the approach of photothermal cancellation, to observe visible photothermal effects the detuning scans had to be slowed down to a greater extent, making the scans more vulnerable to acoustic noise. The resulting errors in fitting parameters from single traces are extremely large. Filtering was considered but not possible due to the similar time-scales between the noises and the cavity response. Instead of fitting single raw traces, we take their average to smooth out the stammering wobbles caused by external noise. Defining the multiple-trace alignment with a fixed reference point to take the average, however, is not straightforward. The best fit relies on the choice of the reference point. We define a weighted averaging method, which takes into account the misfit caused by alignment approach. We calculated the averages of single raw traces by aligning them at the points of \SI{10}{\percent}, \SI{20}{\percent}, \ldots, \SI{100}{\percent} of their maxima. With those averages as data samples, the weights are considered to be inversely proportional to the corresponding errors in the parameter fits. Here, we aligned the traces at different percentage of the maxima on both left- and right-hand sides of the resonance, resulting in the weights $w_{i}$ where $i=1,2,\ldots,19$.

\begin{figure}[h!] \centering
\includegraphics[width=1\columnwidth]{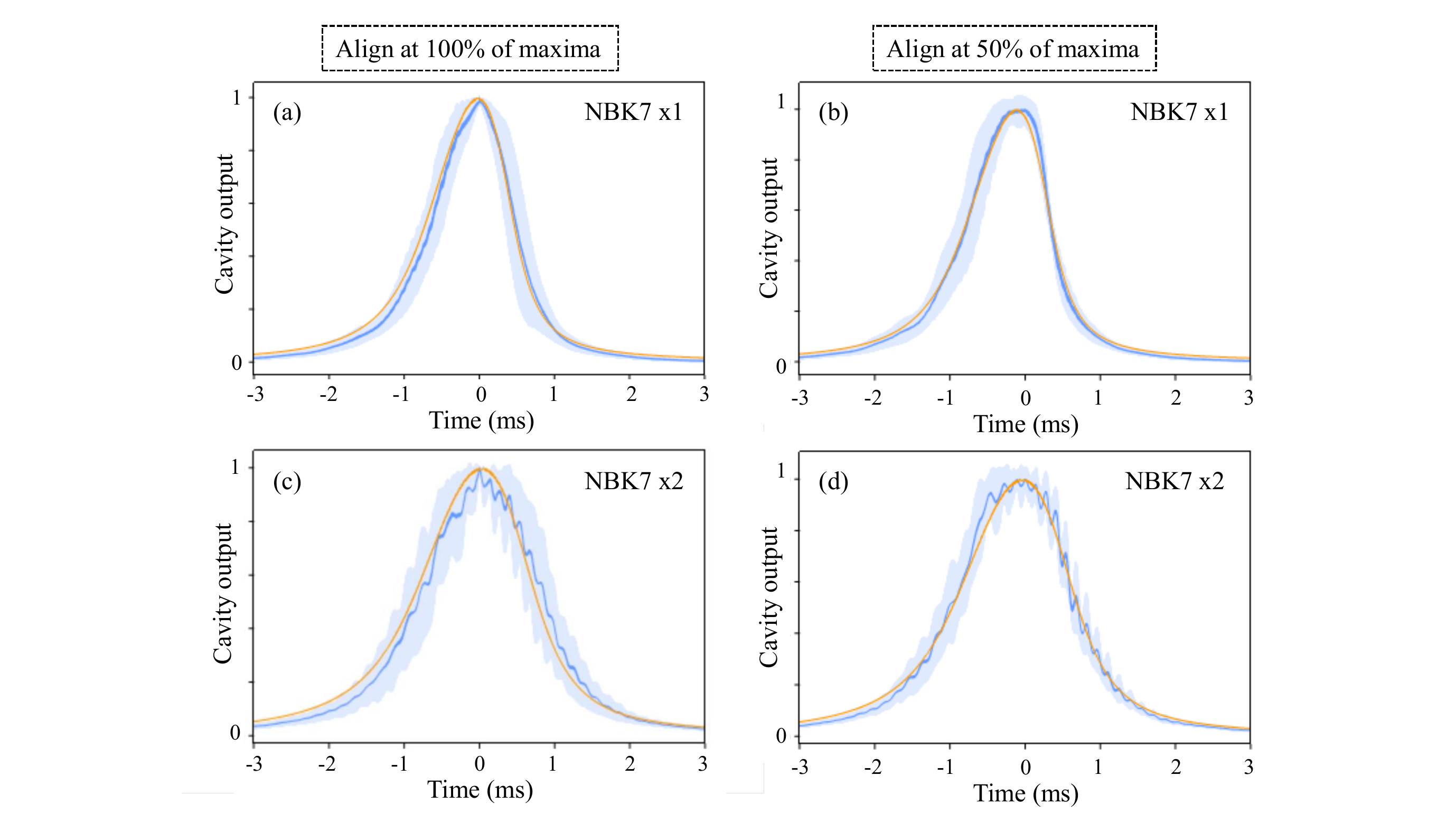} 
\caption{Representative plots of cavity output for (a)-(b) N-BK7 single and (c)-(d) N-BK7 double aligned at 100\% and 50\% of the maxima. The darker blue curves are experimental average. The lighter blue boundaries are given by single traces. The orange curves are theoretical using parameters for N-BK7$\times$1 and N-BK7$\times$2 in Table~\ref{tab: parameters}.} 
\label{fig: SI_2}
\end{figure}

Figure~\ref{fig: SI_2} shows representative plots of the averaged trace at certain alignments for the cavity with single and double N-BK7 wafer windows. The fits are of acceptable confidence. It is worth recalling that the data for the double N-BK7 wafer window is more noisy than that for a single wafer window.

\begin{figure}[h!] \centering
\includegraphics[width=1\columnwidth]{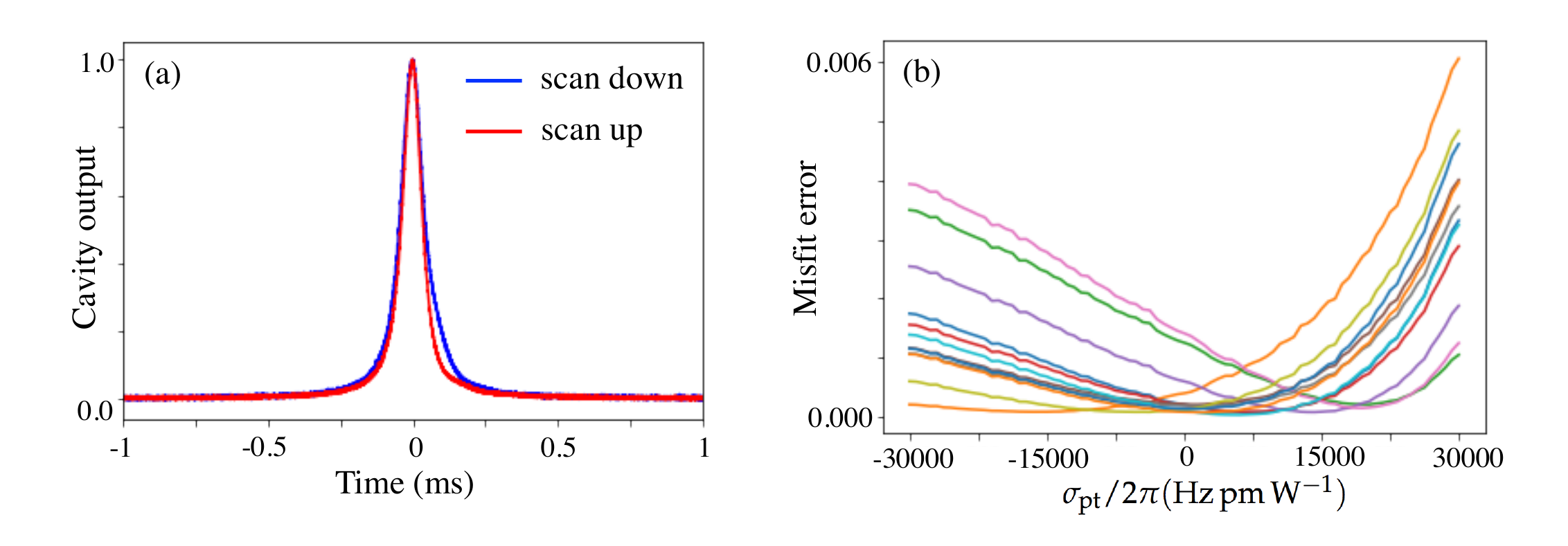} 
\caption{Cavity response and fit error of $4\times\SI{0.22}{\milli\metre}$ N-BK7 wafer windows. (a) Cavity outputs at the downwards and upwards scans. (b) Misfit error versus $\sigma_{\rm pt}$ at different alignments.} 
\label{fig: SI_3}
\end{figure}

In the case of quadruple stack of N-BK7 wafer windows, the cavity response during downwards and upwards scans is nearly symmetric (see Fig.~\ref{fig: SI_3}(a)). This indicates that the photothermal effect is close to zero. We used the same averaging method as the other N-BK7 wafer windows to process the data. The fits turned out to be of low confidence where the best fits for $\sigma_{\rm pt}$ at different alignments spread largely from -16500 up to 19500. Different alignments gave low misfit errors over a large range of $\sigma_{\rm pt}$. In this scenario, the resulting average is not a fair guide to photothermal parameters. But as shown in Fig.~\ref{fig: SI_3}(b), it is fair to conclude that most low misfit errors populate at the values where $\sigma_{\rm pt}$ is near zero. In future work, a better refined photothermal model, which consider factors like separate photothermal effects with individual time constants, and even higher-order corrections to the empirical model of Eq.\ 3, will provide more accurate fits. 


\subsection*{3. Levitation with small photothermal cancellation}
\label{sec: small cancel}
In this section, we show an example of stable levitation achieved through a small negative photothermal parameter ($\sigma_{\rm pt}/2\pi=\SI{-1000}{\hertz\pico\metre\per\watt}$). According to the back-action of relation pressure and photothermal effects shown in Fig.~2f, the effective damping at negative photothermal effects can help the system reach its steady state ($\gamma_{\rm eff}>0$) over a large range of effective detuning. At a weaker negative photothermal regime, the steady state can still be reached but at relatively large effective detunings (Fig.~2f, compare $\beta=10$ pm/W and $\beta=40$ pm/W).

\begin{figure}[t!] \centering
\includegraphics[width=0.6\columnwidth]{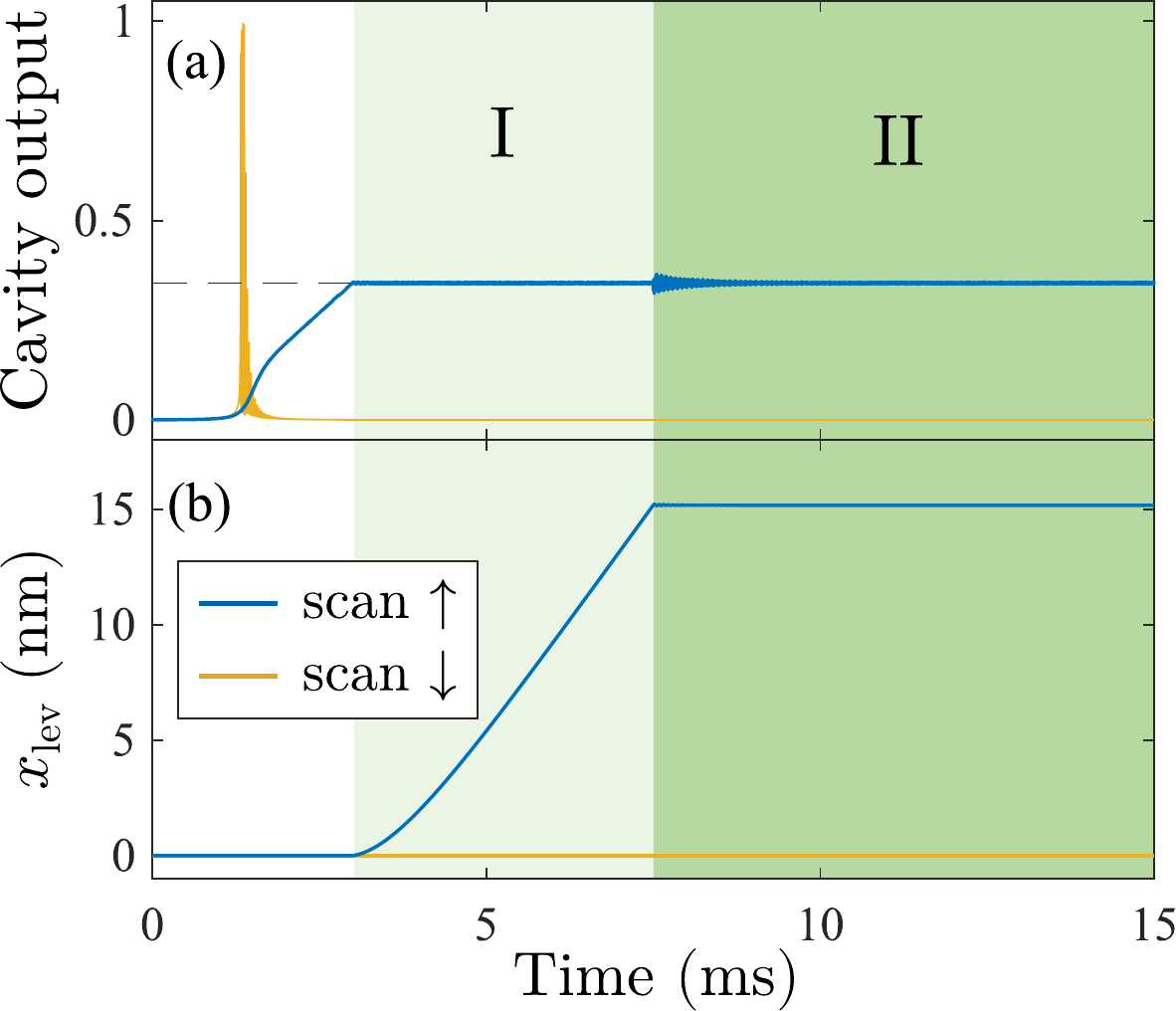} 
\caption{Numerical simulation for a levitation system with a negative photothermal parameter $\sigma_{\rm pt}/2\pi=\SI{-1000}{\hertz\pico\metre\per\watt}$. (a) The cavity outputs, which are normalised to the resonance transmissions. (b) The free-standing mirror's centre-of-mass degree of freedom, $x_{\rm lev}$. The simulations are obtained in presence of a linear detuning scan by the input mirror at the bottom, with direction as indicated by the legend. In all panels, the lighter green shaded area indicates where the cavity power exceeds the threshold for levitation. On the right-hand side, the differently shaded areas indicate: I) lifting of the mirror during the scan; II) sustaining levitation of the mirror at the height when the scan is stopped at \SI{7.5}{\milli\second}.} 
\label{fig: SI_4}
\end{figure}

Fig.~\ref{fig: SI_4} shows the feasibility of stable levitation when $\sigma_{\rm pt}/2\pi=\SI{-1000}{\hertz\pico\metre\per\watt}$. In order to reach levitation at a larger effective detuning, a larger input power $P_{\rm in} = 16$W is applied. Consistently, the levitation level normalised to the cavity resonance is lower than the one in Fig.~2c. Similarly, no amplification of oscillations due to parametric gain appears in the system. The weaker photothermal effect makes the cavity reach the power threshold for levitation in a shorter time scale, where the cavity can automatically lock to an off-resonance detuning (first shaded region, lighter green). We stop the scan at the 7.5 ms (second shaded region, darker green). After registering some minor perturbations, the system maintains a sustaining levitation of the mirror at the height ($\sim$15 nm) when the scan is stopped. 

\end{document}